\documentclass[a4paper,usenatbib,twocolumn]{mnras}

\usepackage[T1]{fontenc}
\usepackage{ae,aecompl}

\usepackage{graphicx}	
\usepackage{amsmath}	
\usepackage{amssymb}	

\title[Magnetic reconnection with IC cooling]
{Particle acceleration in relativistic magnetic reconnection with strong inverse-Compton cooling in pair plasmas}

\author[G.~R.~Werner et al]{
Gregory~R.~Werner,$^{1}$\thanks{E-mail: greg.werner@colorado.edu}
Alexander~A.~Philippov,$^{2}$\thanks{Einstein Fellow}
Dmitri~A.~Uzdensky,$^{1}$
\\
$^{1}$Center for Integrated Plasma Studies, Physics Department, 
   390 UCB, University of Colorado, Boulder, CO 80309, USA\\
$^{2}$Astronomy Department, University of California, Berkeley, 601 Campbell Hall, Berkeley, CA 94720, USA \\
}

\pubyear{2018}

\begin{document}
\label{firstpage}
\pagerange{\pageref{firstpage}--\pageref{lastpage}}
\maketitle

\begin{abstract}
Particle-in-cell (PIC) simulations have shown that relativistic collisionless magnetic reconnection drives nonthermal particle acceleration (NTPA), potentially explaining high-energy (X-ray/$\gamma$-ray) synchrotron and/or inverse Compton (IC) radiation observed from various astrophysical sources. The radiation back-reaction force on radiating particles has been neglected in most of these simulations, even though radiative cooling considerably alters particle dynamics in many astrophysical environments where reconnection may be important. We present a radiative PIC study examining the effects of external IC cooling on the basic dynamics, NTPA, and radiative signatures of relativistic reconnection in pair plasmas. We find that, while the reconnection rate and overall dynamics are basically unchanged, IC cooling significantly influences NTPA: the particle spectra still show a hard power law (index $\geq -2$) as in nonradiative reconnection, but transition to a steeper power law that extends to a cooling-dependent cutoff. The steep power-law index fluctuates in time between roughly $-$3 and $-$5. The time-integrated photon spectra display corresponding power laws with indices $\approx -0.5$ and $\approx -1.1$, similar to those observed in hard X-ray spectra of accreting black holes.
\end{abstract}





\section{Introduction}
\label{sec-intro}

Spectacular high-energy flares in various astrophysical sources are often believed to be powered by magnetic reconnection -- a fundamental plasma process of rapid magnetic field reorganization accompanied by a violent release of magnetic energy and its conversion to plasma energy \cite[e.g.,][]{Zweibel_Yamada-2009}. 
In many systems, reconnection occurs in the relativistic regime, where magnetic energy density exceeds the total (including rest-mass) energy density of the plasma \citep{Lyutikov_Uzdensky-2003, Lyubarsky-2005},
generating relativistic flows, heating the plasma to relativistic temperatures, and driving ultrarelativistic nonthermal particle acceleration (NTPA) \citep{Hoshino_Lyubarsky-2012, Kagan_etal-2015}.
Due to its broad astrophysical relevance, relativistic reconnection has been studied extensively, including via particle-in-cell (PIC) simulations. So far, most of these studies have ignored any radiative aspects of reconnection. However, in many high-energy astrophysical environments, the radiation reaction (which we call {\it radiaction} for short) force on the particles can strongly affect the dynamics, energetics, NTPA, and radiative signatures of reconnection \citep{Uzdensky-2016}.  The two main radiative processes in astrophysical reconnection are synchrotron emission (e.g., in pulsar magnetospheres, \citealt{Lyubarsky-1996, Uzdensky_Spitkovsky-2014, Cerutti_etal-2016, Philippov_Spitkovsky-2018} and pulsar wind nebulae,  \citealt{Uzdensky_etal-2011, Cerutti_etal-2013}) and inverse-Compton (IC) scattering [e.g., in black-hole (BH) accretion disc coronae (ADCe) in X-ray Binaries (XRBs) and active galactic nuclei (AGN), \citealt{Goodman_Uzdensky-2008, Beloborodov-2017}, and also in AGN (e.g., blazar) jets]. 
While several pioneering PIC studies have investigated reconnection with synchrotron cooling 
\citep{Jaroschek_Hoshino-2009, Cerutti_etal-2013, Cerutti_etal-2016, Kagan_etal-2016, Yuan_etal-2016}, IC cooling effects on reconnection have not yet been explored [since submission of this manuscript, \citet{Nalewajko_etal-2018} have also studied IC cooling].

Here we present the first systematic numerical study of relativistic collisionless reconnection with optically-thin external IC cooling.  Using two PIC codes that incorporate the IC radiaction force, {\sc tristan-mp} and \textsc{zeltron}, we show how reconnection-driven
NTPA and resulting radiation signatures change due to the IC drag as accelerated particles scatter off an imposed soft photon bath; we systematically vary the IC cooling via the imposed soft photon density, which represents, e.g., in BH ADCe in the High-Soft (HS) state of~XRBs, the thermal X-rays from the accretion disc.
We limit this first study to $e^+e^-$ pair plasmas, leaving the more XRB-relevant electron-ion case for the future. 
Since previous studies have shown that 2D and 3D PIC simulations of non-radiative pair-plasma reconnection yield very similar NTPA \citep{Werner_Uzdensky-2017}, we use 2D simulations to enable exploration of larger systems. 


\section{Numerical Simulation Setup}
\label{sec-sims}

We use a standard double-periodic box with two relativistic Harris pair-plasma current sheets \citep{Kirk_Skjaeraasen-2003}, plus a uniform background pair plasma of total ($e^-$ and $e^+$) density $n_b$ and temperature $T_b=25 m_e c^2$, reflecting the ambient upstream conditions.
The box dimensions are $L_x \times L_y$ ($L_y=2L_x$), with $x$ parallel to the reconnecting magnetic field~$B_0$ and $y$ perpendicular to the current sheets; $z$ (not simulated) is the initial sheet current direction.
All key parameters are described fully in \citet{Werner_Uzdensky-2017} and briefly in Table~\ref{tab:params} in terms of $B_0$, $n_b$, and $T_b$.
Reconnection is gently kick-started with a small ($10^{-2}$) magnetic field perturbation as in \citet{Werner_Uzdensky-2017}.
We present results for $L_x/\sigma \rho_0 = 320$, well in the large-system regime \citep{Werner_etal-2016} where the high-energy cutoff of the particle spectrum no longer grows linearly with~$L_x$.

\begin{table}
\caption{\label{tab:params}
  Simulation parameters constant over this study.}
\vspace{-2mm}
\begin{tabular}{@{}ll@{}}
 \hline
  nominal gyroradius & $\rho_0=m_e c^2 / eB_0$ \\
  initial guide magnetic field & 
    $B_{gz} = B_0/4$ \\
  ``hot'' magnetization  & 
    $\sigma_h \equiv B_0^2 / (16\pi n_b T_b) = 100$ \\
  ``cold'' magnetization & 
    $\sigma \equiv B_0^2 / (4\pi n_b m_e c^2) = 10^4$ \\
  peak Harris layer density & 
    $n_{d0} = \eta n_b = 5 n_b$ \\
  Harris layer drift velocity & 
    $\pm \beta_d c \hat{\bf z} = \pm 0.3 c\hat{\bf z} $ \\
  Harris layer (comoving) temp. &
    $T_d/m_e c^2 = \gamma_d \sigma /2\eta = 1050$ \\
  Harris layer half-thickness & 
    $\delta  = 
      \sigma \rho_0 / (\eta \beta_d) = 0.67 \sigma \rho_0
      $
      \\
  cell size & 
    $\Delta x=\Delta y = \sigma \rho_0/24$ \\
  total macroparticles per cell & 
    10 \\
  time step & 
    $\Delta t=0.45\Delta x/c$ 
  \\
  simulation time & 
    $T=5L_x/c$ \\
    \hline
 \end{tabular}
\end{table}

Our codes {\sc  tristan-mp} \citep{Spitkovsky-2005} and {\sc zeltron} \citep{Cerutti_etal-2013} use standard PIC algorithms, explicitly evolving Maxwell's equations on a grid with currents self-consistently calculated from particles moving via the Lorentz force.  In addition, they include the back-reaction force on particles emitting IC radiation \citep{Tamburini_etal-2010}.

The IC drag force, felt by an electron or positron with velocity $c\boldsymbol{\beta}$ and energy $\gamma m_e c^2$ as it upscatters photons from an isotropic radiation bath of energy density~$U_{\rm ph}$, is 
$
\boldsymbol{f}_{\rm IC} = -
(4/3) \sigma_T
U_{\rm{ph}}\gamma^2 \boldsymbol{\beta} \, , 
$
where $\sigma_T=8\pi e^4/(3m_e^2 c^4)$.
Balancing $\boldsymbol{f}_{\rm IC}$ against the accelerating force of the reconnection electric field $E=\beta_{\rm rec} B_0$, and taking $\beta_{\rm rec}\sim 0.1$ for relativistic reconnection, yields the radiaction limit: 
\begin{eqnarray}
  \gamma \; \lesssim \; \gamma_{\rm rad} &\equiv & 
  \sqrt{3(0.1) e B_0 /(4\sigma_T U_{\rm{ph}})} \, .
\end{eqnarray}
IC cooling, controlled by $U_{\rm ph}$, is
conveniently quantified by~$\gamma_{\rm rad}$, since
$\left| \boldsymbol{f}_{{\rm IC}} \right| \approx 0.1 e B_0\left( \gamma / \gamma_{\rm rad}\right)^2$ for $|\boldsymbol{\beta}| \approx 1$.

We ran seven simulations differing only in $U_{\rm ph}$: $\gamma_{\rm rad}/\sigma$\allowbreak=1,\allowbreak 2,\allowbreak 4,\allowbreak 6,8,16, and $\infty$ (no radiaction) -- a wide-ranging exploration, since $|\boldsymbol{f}_{\rm IC}|\sim \gamma_{\rm rad}^{-2}$.
Stronger radiaction, $\gamma_{\rm rad} \ll \sigma$, would (for $\sigma_h=100$) cool the upstream plasma before it reaches the current layer, causing the upstream parameters to vary in time and changing the nature of the problem.


\section{Results}
\label{sec-results}


\label{subsec-dynamics}

\begin{figure*}
\begin{center}
\includegraphics[height=4.2cm,trim={0cm 3mm 0mm 4mm},clip=true]{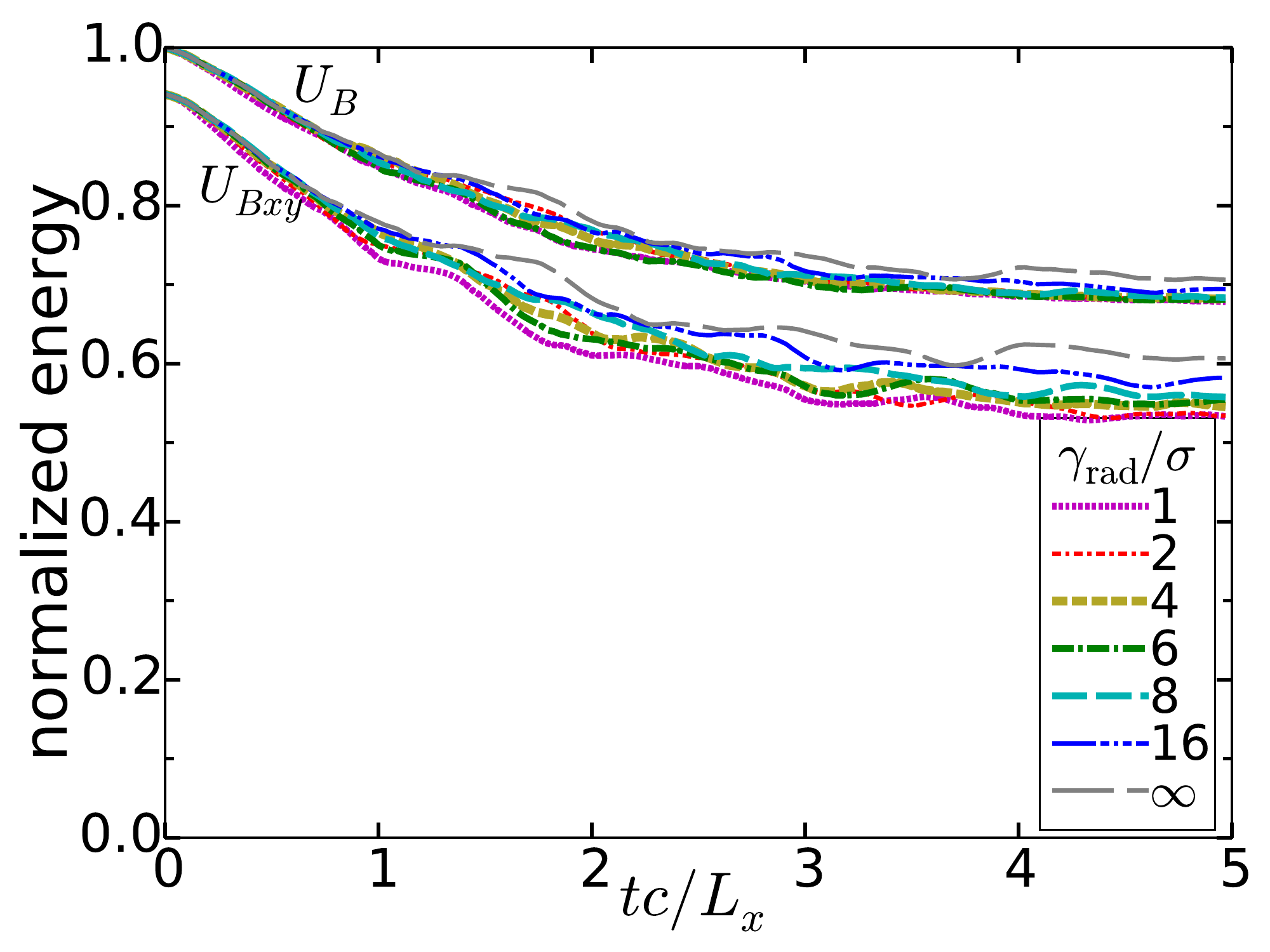}%
\hfill 
\includegraphics[height=4.2cm,trim=0 3mm 0 3mm,clip=true]{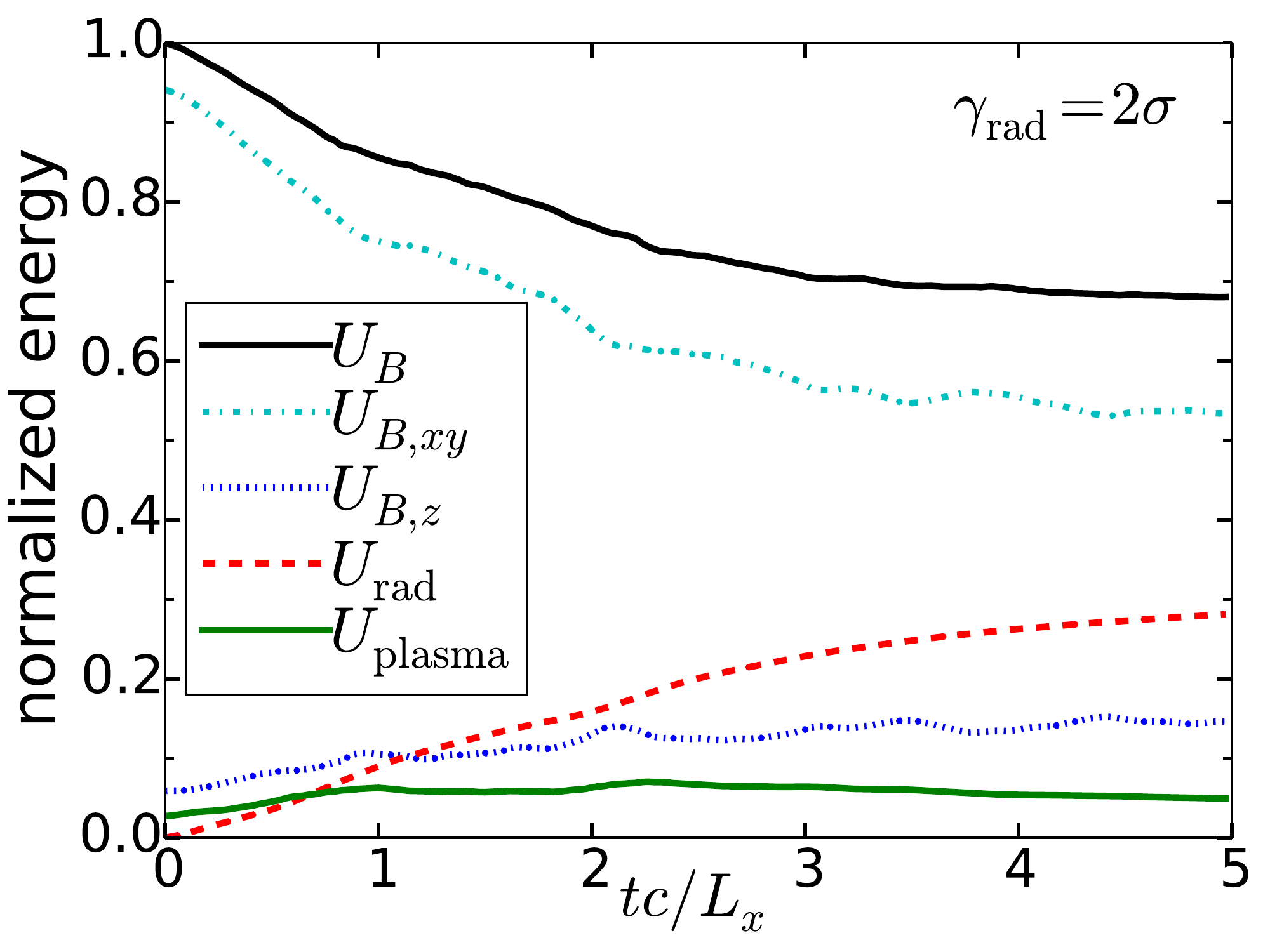}%
\hfill
\includegraphics[height=4.2cm,trim=0cm 3mm 0 3mm,clip=true]{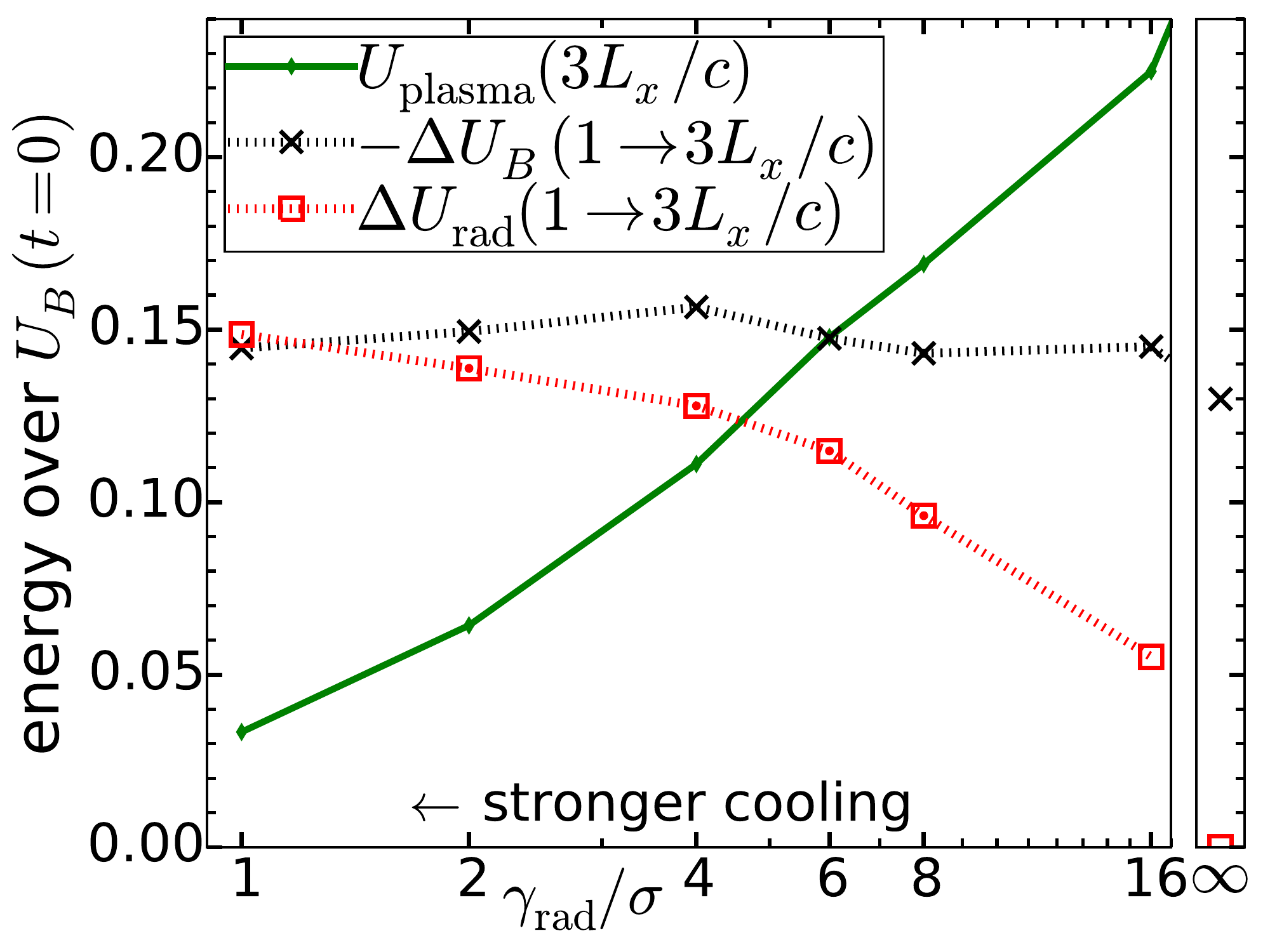}%
\makebox[0cm][l]{\hspace{-\textwidth}\hspace{8mm}\raisebox{4.85mm}{\includegraphics*[width=2.7cm,trim=0 5mm 4mm 2mm,clip=true]{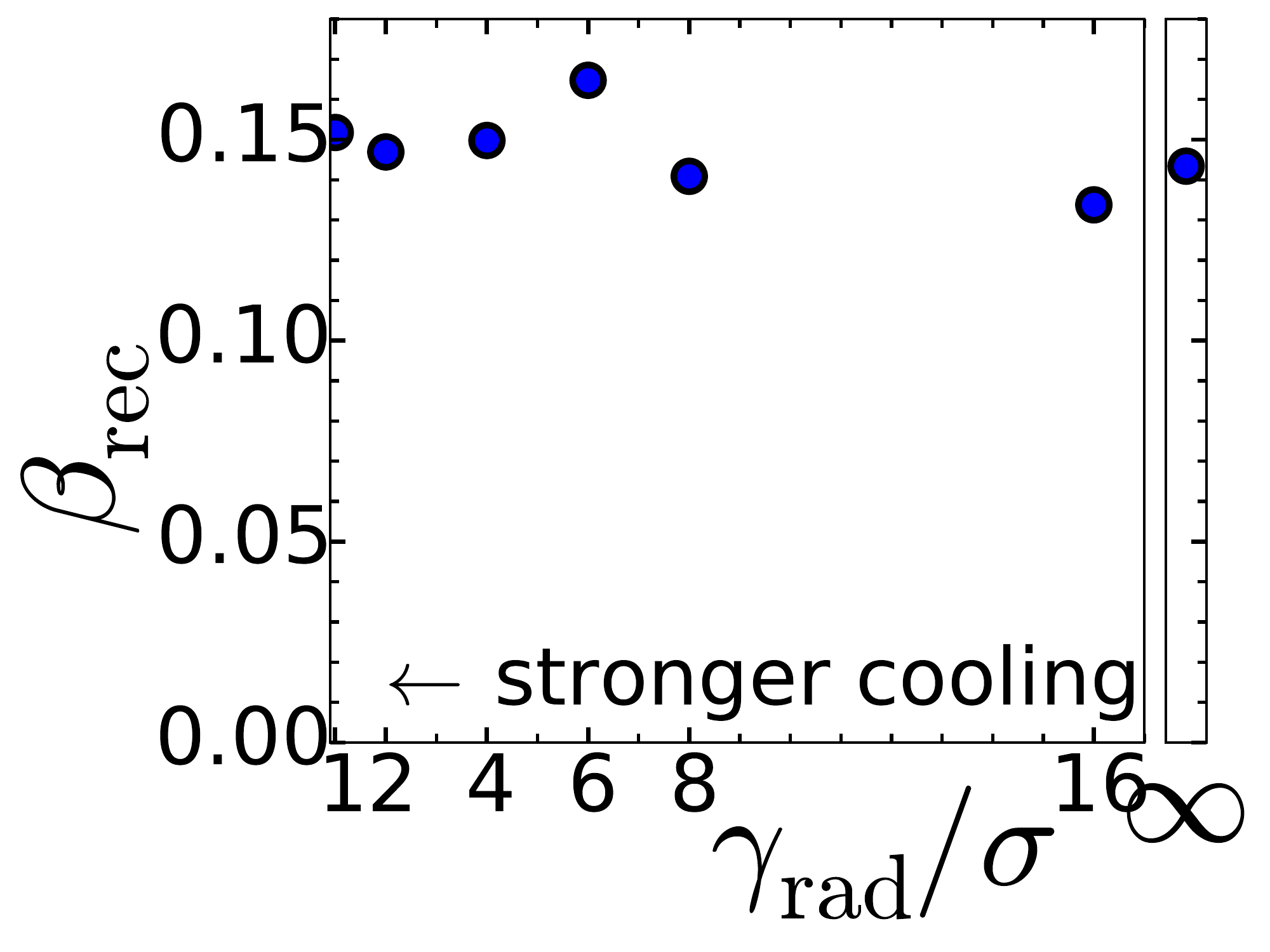}}}%
\makebox[0cm][l]{\hspace{-\textwidth}\raisebox{1mm}{(a)}}%
\makebox[0cm][l]{\hspace{-11.7cm}\raisebox{1mm}{(b)}}%
\makebox[0cm][l]{\hspace{-5.8cm}\raisebox{1mm}{(c)}}%

\caption{
\label{fig:energy}
(a) The total and transverse magnetic energy $U_B(t)$ and $U_{B,xy}(t)$ versus time are nearly independent of radiative cooling strength (given by $\gamma_{\rm rad}$), and, inset, the normalized reconnection rate is also independent of~$\gamma_{\rm rad}$.
(b) For a strongly-cooled simulation, $\gamma_{\rm rad}=2\sigma$, the magnetic energy $U_B(t)$ is similar to weakly-cooled cases, as is the sum of particle $U_{\rm plasma}$ and radiated $U_{\rm rad}$ energies; being strongly-cooled, however, $U_{\rm plasma}(t)$ quickly saturates, after which any particle energy gains are promptly radiated away.
(c) The magnetic energy dissipated between~1 and 3$L_x/c$ is independent of cooling strength; for weak cooling, it increases the plasma energy, while for strong cooling it is no sooner given to particles than it is radiated away. All energies are normalized to~$U_B(t=0)$.
}
\end{center}
\end{figure*}

{\bf Reconnection Dynamics and Energetics.}  Reconnection begins as the tearing instability breaks up the current layer into chains of magnetic islands (plasmoids) separated by magnetic X-points. 
Over time, plasmoids merge into larger ones, while secondary current sheets between them succumb to secondary tearing, yielding a hierarchical structure of X-points and plasmoids \citep[]{Bhattacharjee_etal-2009,Uzdensky_etal-2010}. 
This familiar picture remains largely unchanged by IC cooling; reconnection continues to perform its most basic function, converting magnetic energy to particle kinetic energy, almost regardless of radiation; radiative cooling merely converts some of that particle energy to radiation.

Although IC cooling strongly affects particles that gain high energies during reconnection, it has little effect on the overall reconnection dynamics and energy conversion, which are evidently 
controlled by the lower-energy particles that, for $\gamma_{\rm rad}\gtrsim \sigma$, experience negligible cooling (cf.~\S\ref{sec-sims}).
Notably, as shown in Fig.~\ref{fig:energy}(a), magnetic energy release proceeds nearly independently of IC cooling and we see no discernible effect of radiaction on the reconnection rate, $\beta_{\rm rec}\sim 0.15$. 
There is, however, a modest effect of radiaction on magnetic dissipation: reconnection with strong radiaction converts slightly more transverse magnetic energy to guide-field energy~$B_z^2/8\pi$. This is because increased cooling reduces plasma pressure in plasmoids, leading to a stronger compression of the guide magnetic field in them.

Although reconnection converts magnetic to particle energy at essentially the same rate, strong radiative cooling ($\gamma_{\rm rad} \lesssim 4\sigma$) causes this energy to be promptly radiated away, maintaining the total kinetic energy $U_{\rm plasma}$ at a nearly constant radiaction-limited level [Fig.~\ref{fig:energy}(b) shows the evolution of magnetic, plasma, and radiated energies for $\gamma_{\rm rad}=2\sigma$].  Therefore, in the strong cooling regime the IC luminosity reaches a universal value ${\rm d}U_{\rm rad}/{\rm d}t \sim 0.1 U_{B,xy}/(L_x/c)$. 

Fig.~\ref{fig:energy}(c) shows how radiaction affects the energy partition between particles and radiation (after about $5 L_x/c$, when reconnection is long over).  With strong cooling, particles radiate away their energy even as they are accelerated by reconnection. Weaker cooling allows particles to reach higher energies before radiaction balances the acceleration due to reconnection. In the limit of very weak cooling, particles are accelerated almost as without cooling, slowly radiating away energy long after exiting the reconnection region.


\label{subsec-NTPA}

{\bf Particle Acceleration.}  Recent PIC simulations have clearly shown NTPA driven by relativistic reconnection.  
Reconnection accelerates a large fraction of particles to high energies $\gamma \gtrsim \sigma$,  
yielding nonthermal high-energy spectra characterized by a power-law index (slope) $p$ and a high-energy cutoff~$\gamma_c$. IC drag, which scales as~$\gamma^2$, can, however, suppress NTPA at highest energies. 

With radiaction, high-energy particle spectra $f(\gamma)$ vary more in time and display more complicated forms than the familiar single power law with a high-energy cutoff. 
To measure spectral slopes and cutoffs, we calculated the local slope $p(\gamma)= -{\rm d} \ln f/{\rm d}\ln \gamma$ and searched for the longest stretch over which $p(\gamma)$ varied within $\pm 10$ per cent; the next-longest power-law stretch was also identified.  We counted only power laws stretching over a factor $\geq 2$ in $\gamma$, and identified the power-law index $p$ as the median~$p(\gamma)$.  The high-energy cutoff $\gamma_c$ was defined by $f(\gamma_c) = {\rm e}^{-1} A\gamma_c^{-p}$, with $A\gamma^{-p}$ being the best fit for $f(\gamma)$ over the power-law stretch.

\begin{figure*}
\begin{center}
\includegraphics*[height=4.4cm,trim=0mm 2mm 0 0mm,clip=true]{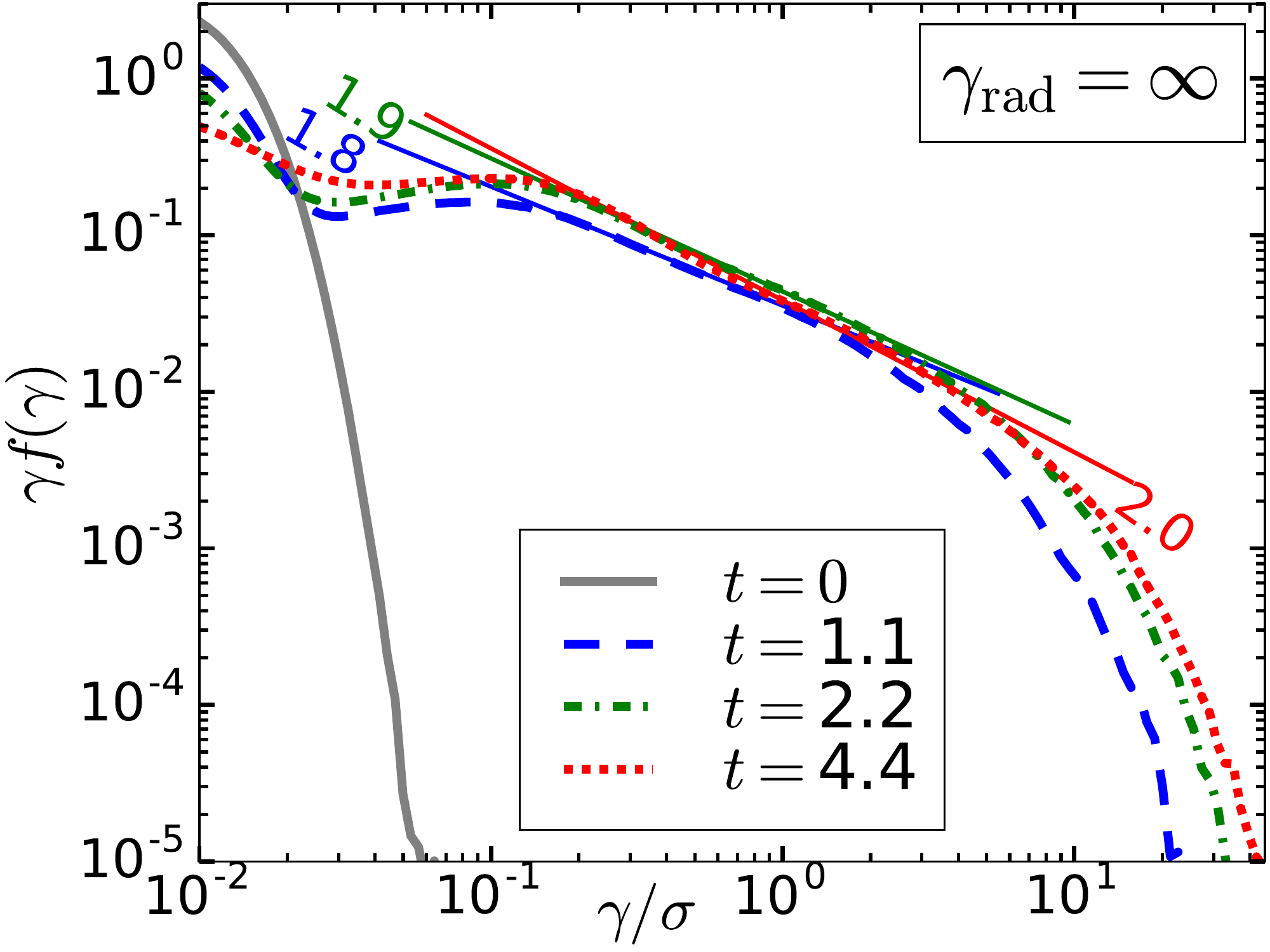}%
\hfill
\includegraphics*[height=4.4cm,trim={24mm 2mm 0 0mm},clip=true]{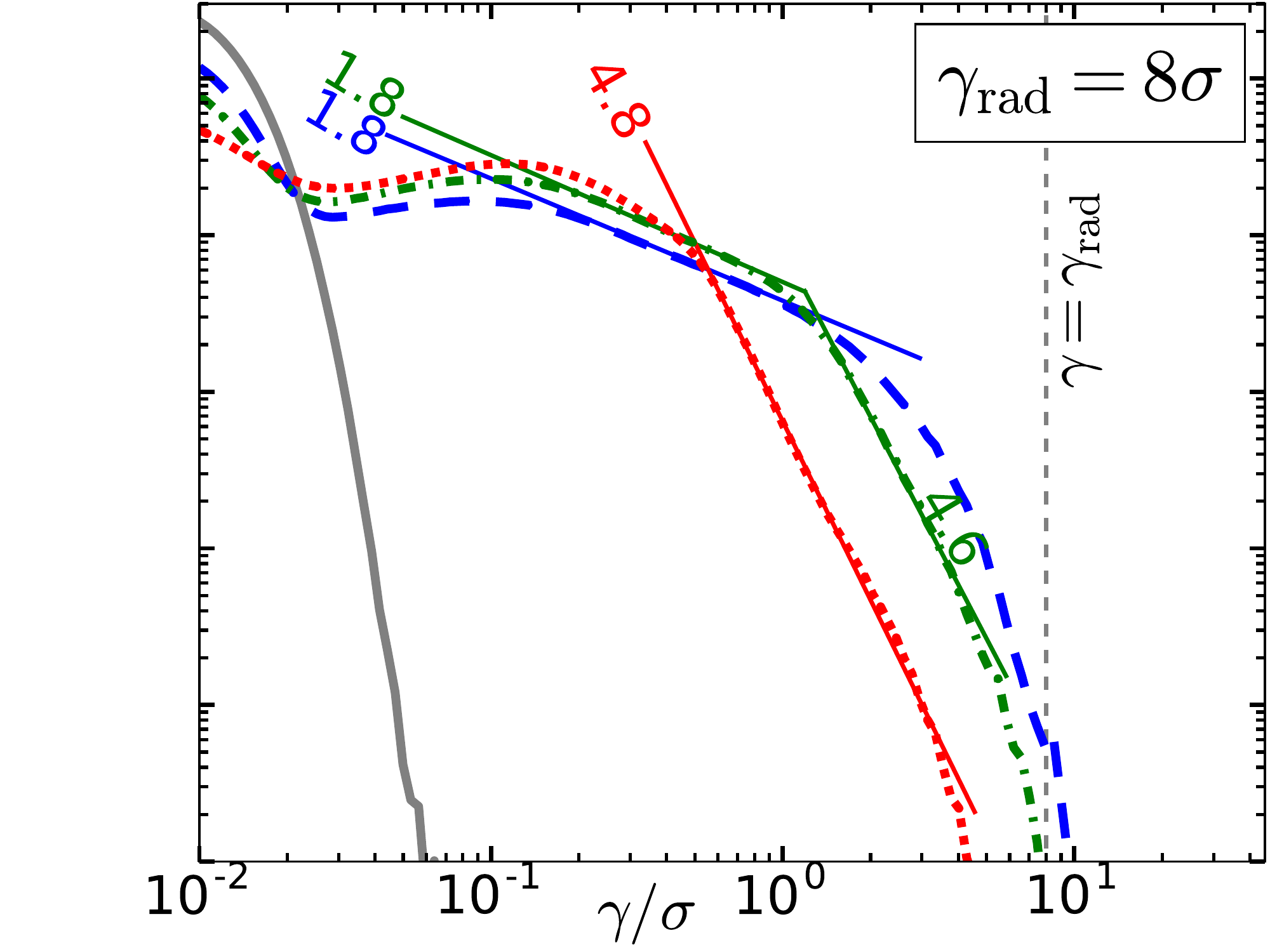}%
\hfill
\includegraphics*[height=4.4cm,trim=24mm 2mm 0 0mm,clip=true]{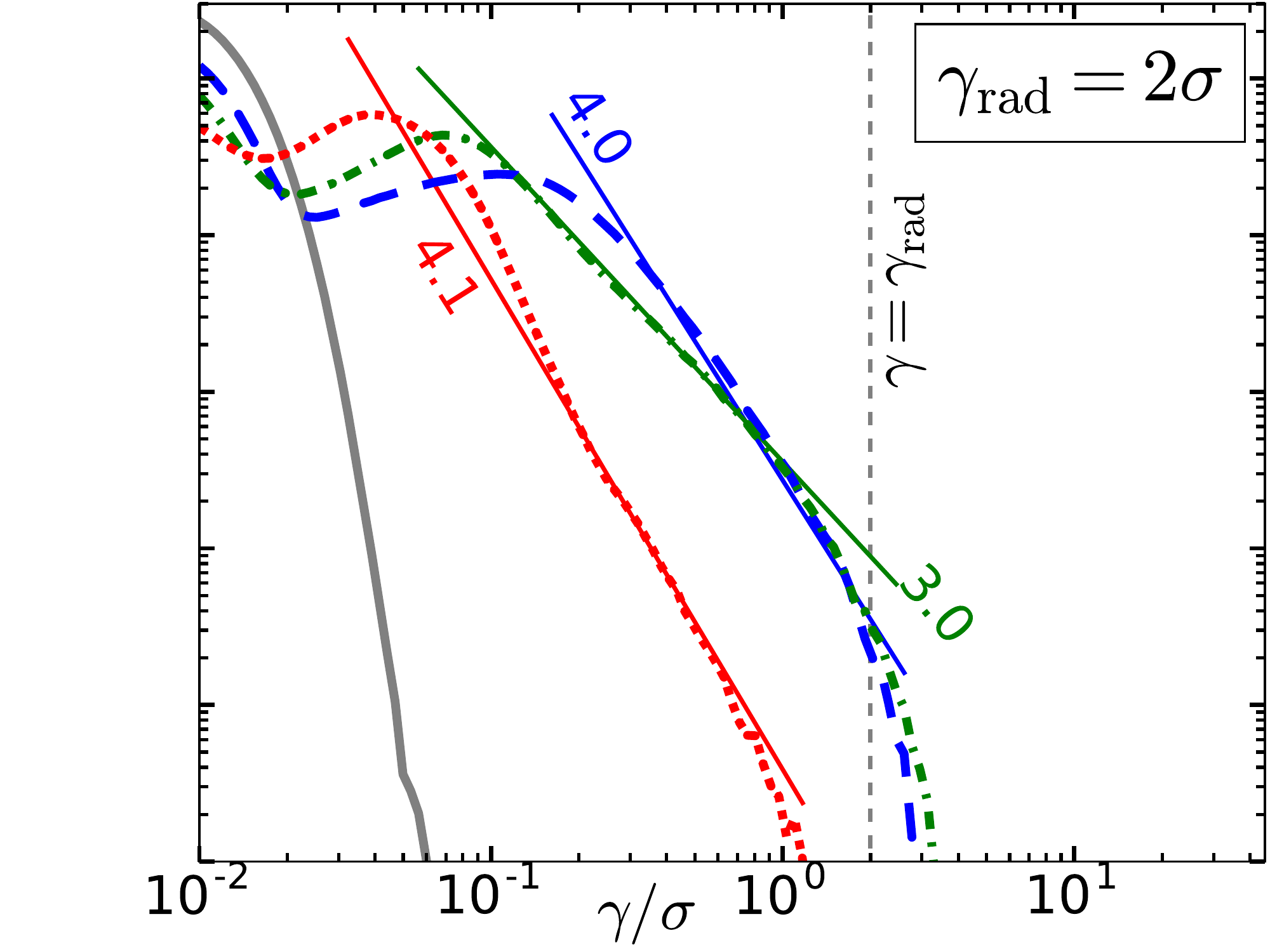}%
\makebox[0mm][l]{\hspace{-\textwidth}(a)}%
\makebox[0mm][l]{\hspace{-\textwidth}\hspace{6.1cm}(b)}%
\makebox[0mm][l]{\hspace{-\textwidth}\hspace{12.cm}(c)}%
\caption{
\label{fig:spectra}
For (a) no cooling, (b) moderate, and (c) strong cooling ($\gamma_{\rm rad}/\sigma=\infty, 8, 2$), the  compensated energy spectra $\gamma f(\gamma)$ are shown at 4 times (0, 1.1, 2.2, 4.4$L_x/c$), with fits $f(\gamma)\sim \gamma^{-p}$ marked with the index~$p$.  For $\gamma_{\rm rad}=8\sigma$, two power laws are visible at $t=2.2L_x/c$.
}
\end{center}
\end{figure*}

\begin{figure*}
\begin{center}
\includegraphics[height=4.3cm,trim=0 2mm 0 0mm,clip=true]{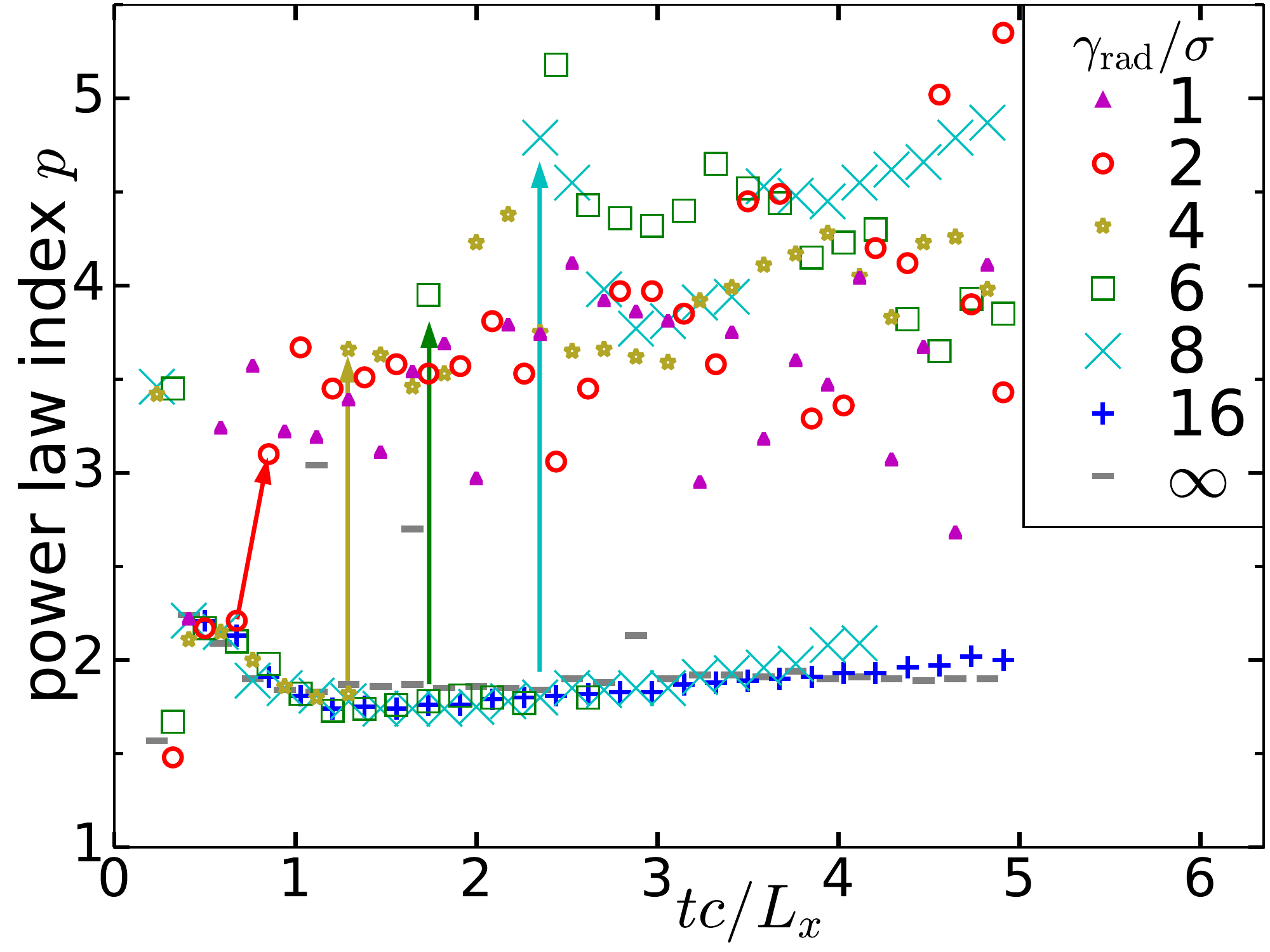}%
\hfill
\includegraphics[height=4.3cm,trim={0cm 2mm 0 0mm},clip=true]{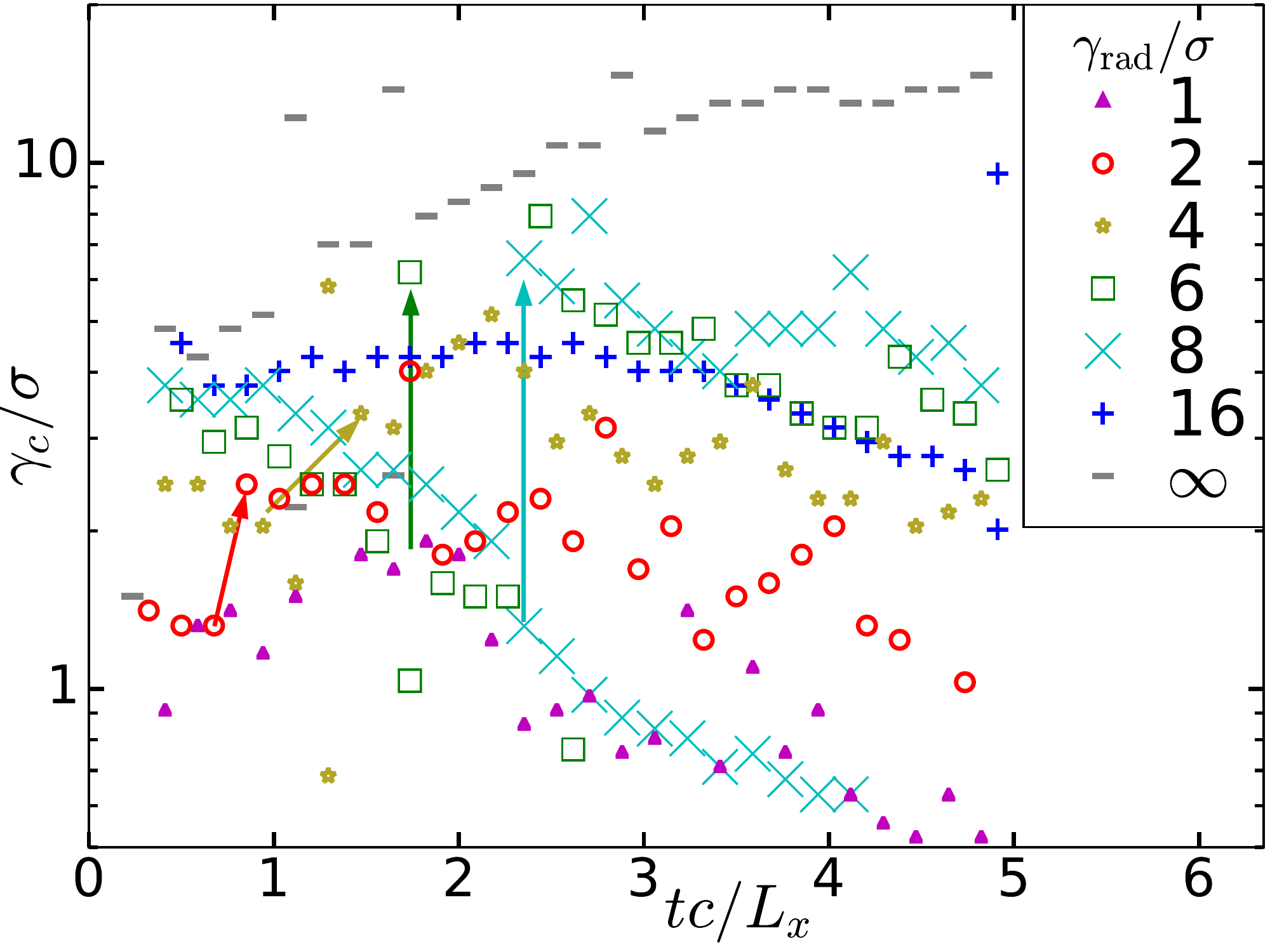}%
\hfill
\includegraphics[height=4.3cm,trim=0cm 0mm 0 0mm,clip=true]{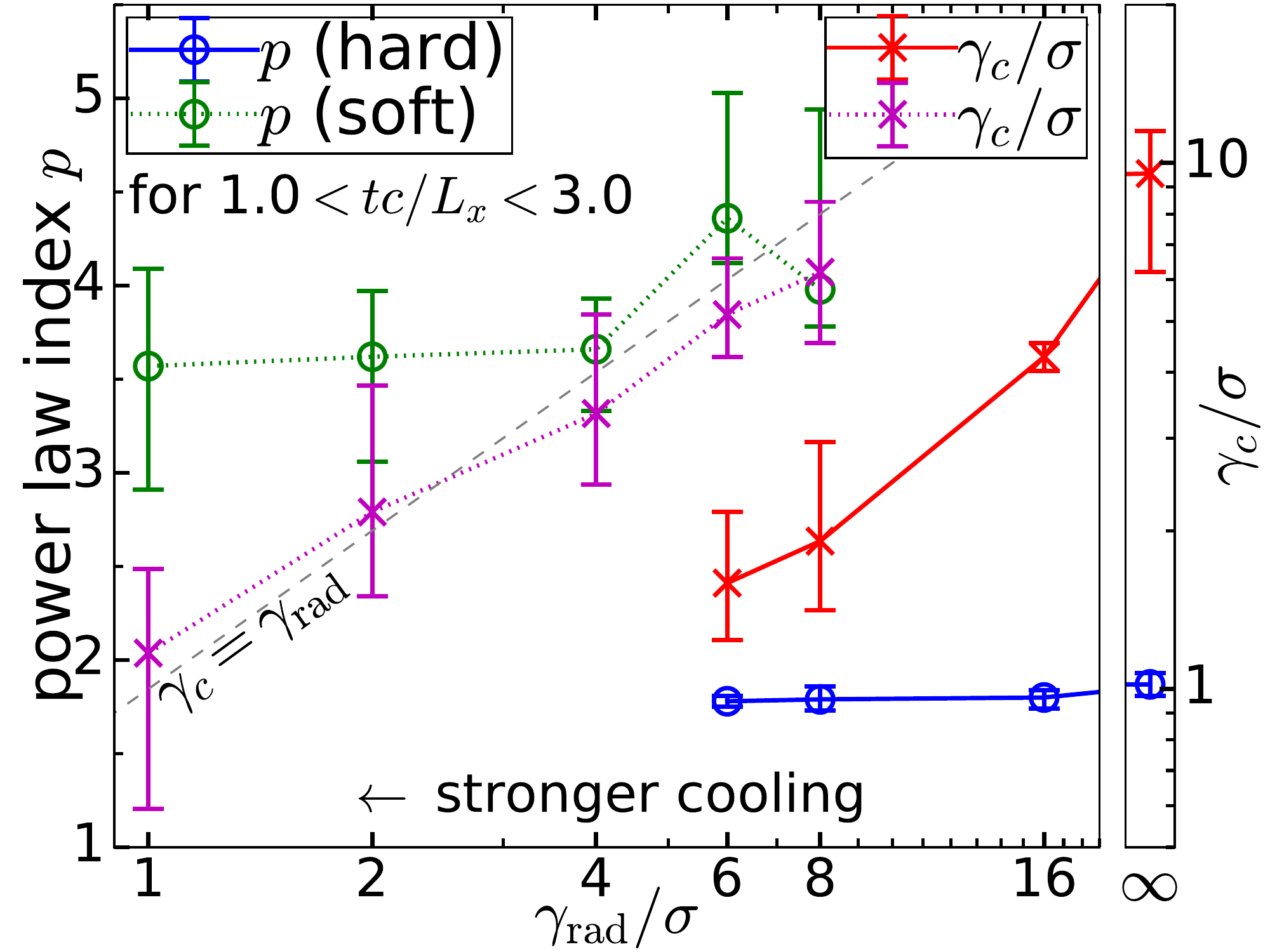}%
\makebox[0mm][l]{\hspace{-\textwidth}(a)}%
\makebox[0mm][l]{\hspace{-\textwidth}\hspace{5.9cm}(b)}%
\makebox[0mm][l]{\hspace{-\textwidth}\hspace{12.cm}(c)}%
\caption{
\label{fig:plaws}
Power-law (a) indices and (b) high-energy cutoffs [where $f(\gamma_c)$ is a factor $e$ below the power-law fit] versus time, for several cooling strengths; if hard and soft power laws appear in the same spectrum, we show $p$ and $\gamma_c$ for both, connected by an arrow the first time it occurs. (c) The median power-law indices (over hard, $p\leq 2.6$, and soft, $p>2.6$, power laws) and corresponding cutoffs for spectra during $1 < tc/L_x < 3$, with error bars containing the middle 80 per cent of measurements in this time interval.  
}

\end{center}
\end{figure*}

Figure~\ref{fig:spectra} shows $f(\gamma)$ at times $tc/L_x=0$, 1.1, 2.2, and~4.4 for (a) no cooling, (b) moderate cooling, and (c) strong cooling, with power-law fits $f(\gamma)\sim \gamma^{-p}$.  Figure~\ref{fig:plaws}(a--b) then shows, for all cooling strengths, the complete time-evolution of the fitted power-law indices $p$ and cutoffs $\gamma_c$.  In the nonradiative case ($\gamma_{\rm rad}=\infty$) a hard power law quickly develops and remains roughly constant in time (aside from overall amplitude) with index $p_h\approx 1.9$.  This agrees with previous studies \citep[e.g.,][]{Sironi_Spitkovsky-2014,Guo_etal-2014,Werner_etal-2016} after accounting for the steepening effect of guide field \citep{Werner_Uzdensky-2017} -- e.g., we find $p_h\approx 1.6$ for $B_{gz}=0.05 B_0$.  The high-energy cutoff rises rapidly to $\gamma_c \simeq 4\sigma$ \citep[as in][]{Werner_etal-2016} and then slowly to $\simeq 10\sigma$ \citep[cf.][]{Petropoulou_Sironi-2018arxiv} before reconnection ends around $t\simeq 3L_x/c$.
With weak cooling ($\gamma_{\rm rad}=16\sigma$) a similar hard power law forms up to (but not beyond) $\gamma_c\simeq 4\sigma$, steepening slightly with time while the cutoff $\gamma_c$ decreases (as higher-energy particles cool faster).
For moderate cooling ($\gamma_{\rm rad}=6,8\sigma$), a hard power law forms at early times with the uncooled $p_h\approx 1.9$, extending initially to $\gamma_c\simeq 4\sigma$.  Over time, however, cooling steepens the high-energy part of the power law, yielding a broken power law: the same hard slope $p_h\approx 1.9$ at lower energies, and at higher energies a steeper power law with a highly-variable index [$p_s\approx 4.6$ is shown at $t=2.2L_x/c$ in Fig.~\ref{fig:spectra}(b), but Fig.~\ref{fig:plaws}(a) shows that $3 \lesssim p_s \lesssim 5$ at other times].
Over time, cooling shifts the break $\gamma_{\rm br}$ between power laws toward lower energies until the hard power law disappears.
For strong cooling ($\gamma_{\rm rad}=1,2,4\sigma$), the hard power law appears tenuously at the beginning of active reconnection, but is quickly replaced by the soft/steep power law ($p_s\gtrsim 3$) that dominates the remaining time; as particles no longer being accelerated continue to cool, a low-energy ``pile-up'' forms, e.g., at $\gamma\simeq 0.04\sigma$ for $t=4.4 L_x/c$ in Fig.~\ref{fig:spectra}(c).  Although stronger cooling (lower $\gamma_{\rm rad}$) reduces the high-energy cutoff to $\gamma_c\sim \gamma_{\rm rad}$ during active reconnection, the steep power law index $p_s$ varies over time within roughly the same range (3--5) for all cases $\gamma_{\rm rad}\lesssim 8 \sigma$.
Similarly, when visible, $p_h \approx 1.9$ is independent of cooling strength.

In Fig.~\ref{fig:plaws}(c) we summarize this picture for the time of active reconnection, roughly 1--3$L_x/c$.  Independent of $\gamma_{\rm rad}$, the hard power-law index is $p_h\approx 1.9$, and the soft power-law index falls in the range $3<p_s<5$. While steady-state models for radiatively cooled broken power laws predict an increase of $p$ by~1, reconnection-driven NTPA is non-steady, with bursts of efficient acceleration at X-points followed by cooling. Thus, the soft/steep power law varies in time, reaching a minimum slope of $p_{s,\rm min}\approx 3\approx p_h+1$ occasionally and becoming much steeper than $p_h+1$ during uninterrupted cooling episodes.
Figure~\ref{fig:plaws}(c) also shows that the cutoff of the soft/steep power law (when it exists, i.e., for $\gamma_{\rm rad}\leq 6\sigma$) scales as~$\gamma_{\rm rad}$. 
The cutoff of the hard power law (i.e., $\gamma_{\rm br}$) decreases with stronger cooling, and the hard power law becomes almost indiscernible for $\gamma_{\rm rad}\lesssim 4\sigma$.

\label{subsec-radiation}

\begin{figure}
\begin{center}
\includegraphics[height=5.5cm,trim=0 0 0 0]{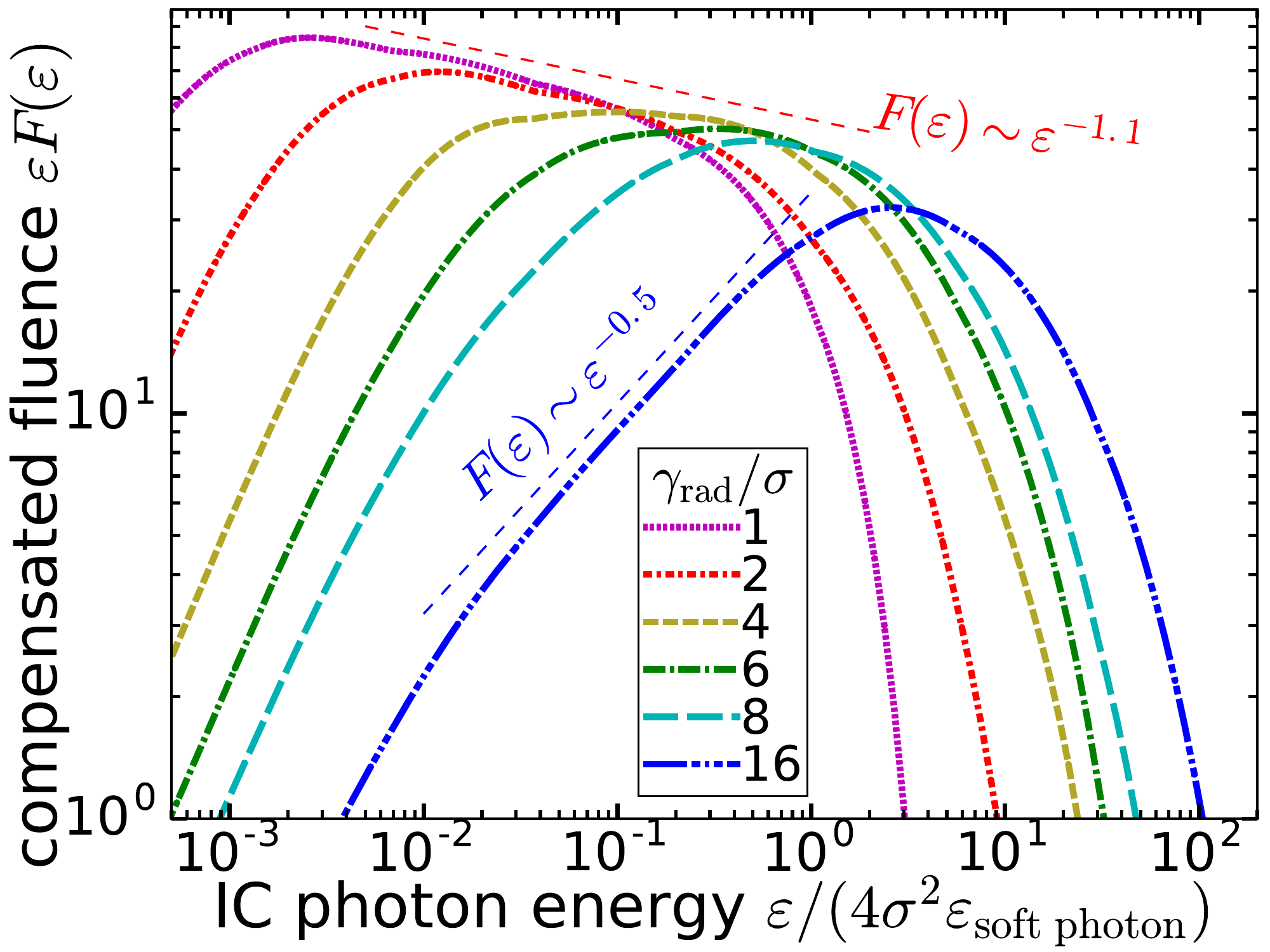}%
\caption{
\label{fig:radiation}
Time-integrated photon spectra (compensated by $\varepsilon$) for different cooling strengths.  (Note: an electron with $\gamma$ upscatters an IC photon to maximum energy $4\gamma^2 \epsilon_{\rm soft\ photon}$, so a $\gamma=\sigma$ electron emits photons of maximum normalized energy~1.)
}
\end{center}
\end{figure}

{\bf Radiation.}  Figure~\ref{fig:radiation} presents IC radiation spectra $F(\epsilon)$ integrated over the simulation time, i.e., over an entire reconnection flare. 
The power-law index of IC radiation emitted by particles with steady-state power-law index $p$ should be $\alpha_{IC}=(p-1)/2$, consistent with our measurements $p_h\approx 1.9$ and $\alpha_{IC}\approx 0.5$ for weak cooling, $\gamma_{\rm rad}=16\sigma$.
For strong cooling ($\gamma_{\rm rad}=2\sigma$), the instantaneous particle and hence photon spectra vary greatly with time. However, periods with harder spectra, $p_s(t) \approx p_{s,\rm min}\approx 3$, should dominate the overall high-energy emission,
i.e., $\alpha_{IC} \approx (p_{s,\rm min}-1)/2 \approx 1$, in agreement with our measured value $\alpha_{IC}\approx 1.1$.

\label{subsec-codes}

\begin{figure}
\begin{center}
\begin{tabular}{l@{}l}
(a)&\includegraphics*[height=5.3cm,trim=1mm 0 0 0,clip=true]{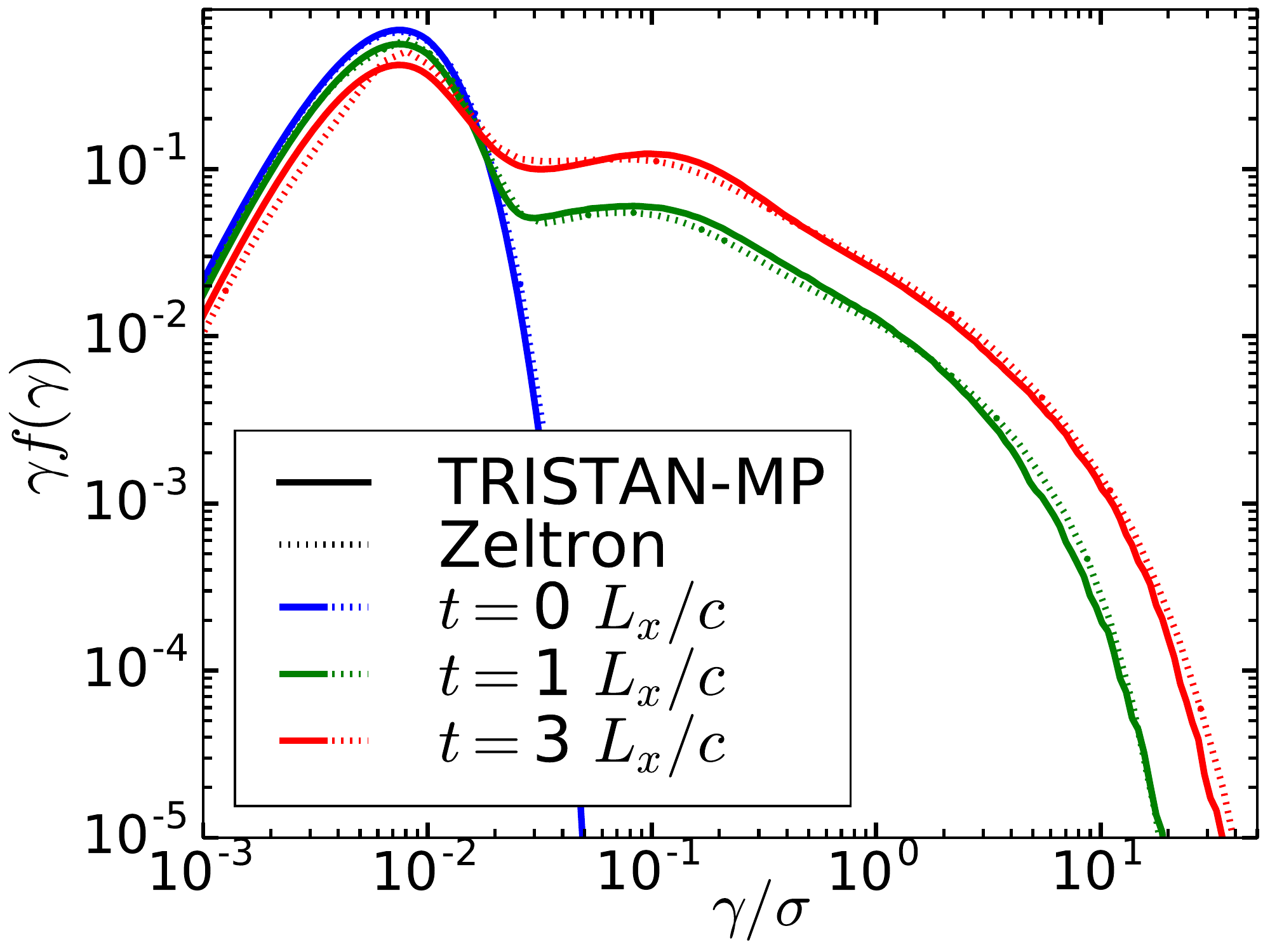}%
\\
(b)&\includegraphics*[height=5.3cm,trim=1mm 0 0 0,clip=true]{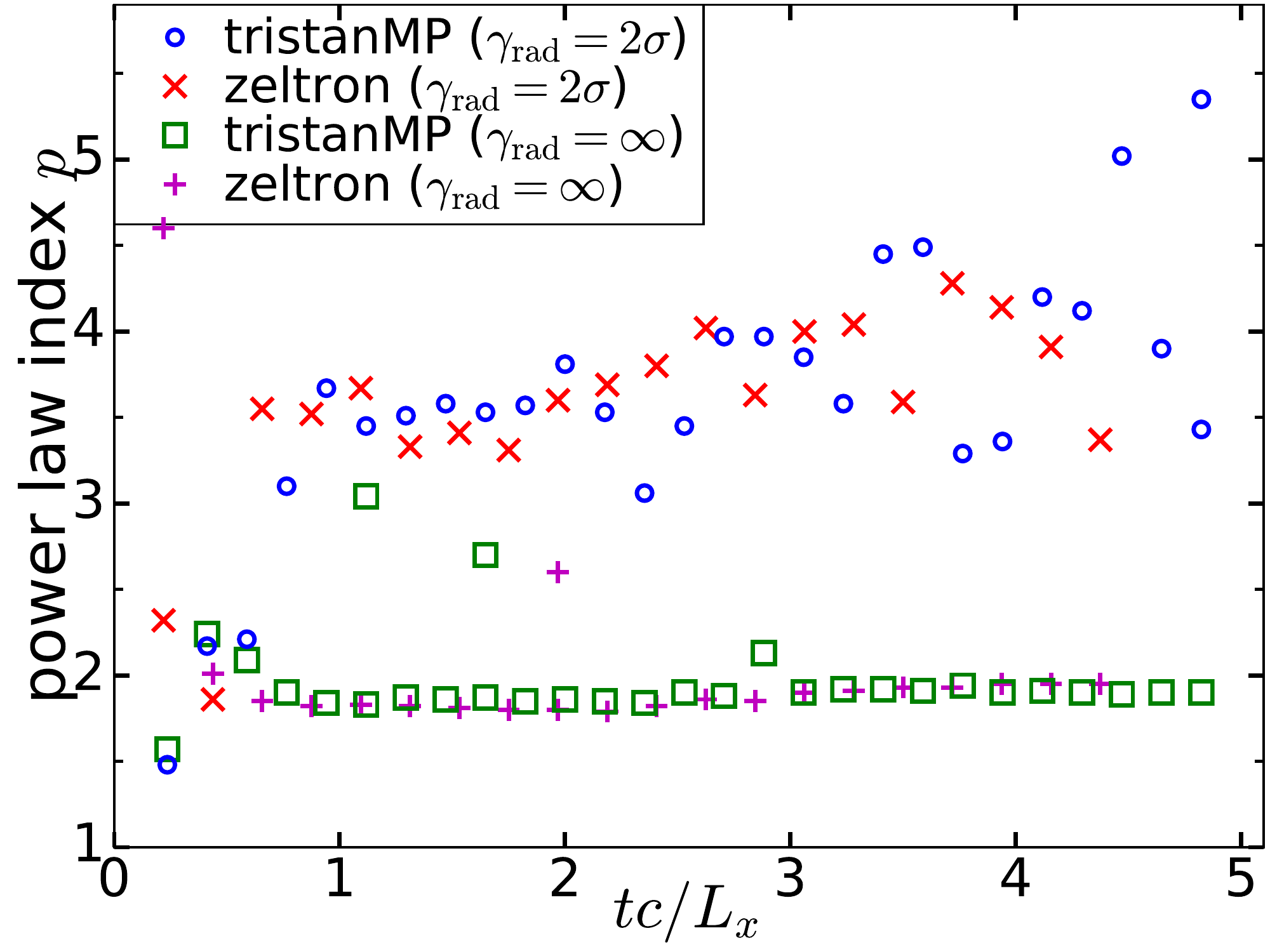}%
\end{tabular}
\caption{
\label{fig:tzCmp}
{\sc tristan-mp} and {\sc zeltron} yield very similar results. 
(a) Particle energy spectra of {\sc tristan-mp} (solid lines) and {\sc zeltron} (dotted) at different times for a nonradiative simulation (stochastic fluctuations prevent sufficiently precise comparison of radiative simulations);
(b) power-law slopes for $\gamma_{\rm rad}=2\sigma$ and~$\infty$.
}
\end{center}
\end{figure}

{\bf Code Comparison.}  {\sc tristan-mp} and {\sc zeltron} implement essentially the same algorithms, including the IC radiation reaction (\S\ref{sec-sims}), but have minor differences, e.g., in charge-conserving current deposition [\citet{Umeda_etal-2003} versus \citet{Esirkepov-2001}].
Despite wide use in astrophysics, the two codes have not yet been systematically compared; fortunately, we find that they produce essentially identical results for radiative reconnection.  
Although noisy, 
the magnetic energy evolution matches closely over long time scales. Crucial to our study of NTPA, the particle spectra are remarkably similar [Fig.~\ref{fig:tzCmp}(a)], agreeing very closely on spectral indices~$p$ [Fig.~\ref{fig:tzCmp}(b)] and cutoffs $\gamma_c$ (not shown) for $\gamma_{\rm rad}=\infty$. For the strongly-radiative case  $\gamma_{\rm rad}=2\sigma$, stochastic time variation prevents precise comparison at any given time, but both codes yield variation within the same range.


\section{Conclusions}
\label{sec-conclusions}

We presented the first systematic numerical study of the effects of the IC radiation reaction (`radiaction') on magnetic reconnection using first-principles PIC simulation.
We found that, even with strong cooling, basic reconnection dynamics, such as the reconnection rate and magnetic energy dissipation, are unchanged.
However, IC cooling strongly affects NTPA and the particle energy spectrum. As a result of radiaction, the high-energy spectrum has, in principle, two power laws: 
for $\gamma<\gamma_{\rm br}$, a hard slope as in nonradiative simulations ($p_h\simeq 1.8$--$2$ for $\sigma_h=100$ and $B_{gz}=B_0/4$, independent of $\gamma_{\rm rad}$), and for $\gamma > \gamma_{\rm br}$ a steeper slope $p_s\gtrsim 3$.  Over time, $\gamma_{\rm br}$ decreases.
For weak radiaction ($\gamma_{\rm rad} \gg 4\sigma$; e.g., $\gamma_{\rm rad}=16\sigma$ for the system size studied here), $\gamma_{\rm br}>4\sigma$ and only the hard power law appears, with high-energy cutoff $\gamma_c \simeq 4\sigma$.  Importantly, even weak radiaction prevents the slow growth of $\gamma_c$ beyond $\simeq 4\sigma$ that occurs without cooling.
As radiaction is increased and $\gamma_{\rm br}$ falls below $4\sigma$, both power laws appear simultaneously for a time, with $\gamma_c \simeq \gamma_{\rm rad}$.  For strong radiaction ($\gamma_{\rm rad} \lesssim 4\sigma$), the hard power law is barely detectable.
Reflecting the bursty nature of plasmoid-dominated reconnection, the steep power-law index $p_s$ fluctuates strongly, roughly within 3--5, but the hard power-law ($p_h$), built up over time, is much steadier.
Thus, $p_s \gtrsim p_h + 1$, with equality expected for radiative steepening of a continuously injected power law $p_h$ subject to IC cooling, and inequality corresponding to further cooling between acceleration episodes.
The IC spectra accordingly have two power laws with slopes of roughly $\alpha_{\rm IC}\approx(p_h-1)/2$ and $\alpha_{\rm IC}\approx (p_{s,\rm min}-1)/2\sim p_h/2$.  Lowering the guide field (e.g., to $B_{gz}=0.05 B_0$) yields very similar results but slightly hardens all spectra, as expected \citep{Werner_Uzdensky-2017}.
While IC and synchrotron radiaction
\mbox{\citep[e.g.,][]{Kagan_etal-2016}} 
yield broadly similar qualitative features, such as spectral steepening, a useful quantitative comparison will require a future synchrotron study that systematically varies radiaction strength as in this study.

The robust dichotomy of nonthermal spectra produced by reconnection with IC cooling has important implications for understanding radiative kinetic plasma processes in astrophysical systems like BH ADCe. 
The spectral indices of IC radiation from our simulations of strongly-cooled relativistic reconnection, $\alpha_{\rm IC} \simeq 1.1$, are close to $\alpha \simeq $1.5 observed in hard X-ray spectra (believed to come from IC scattering of soft disc photons by energetic coronal electrons) in HS and Steep Power Law XRB states, while $\alpha_{\rm IC} \simeq 0.5$ seen in weak-cooling simulations is similar to $\alpha \simeq 0.7$ observed in the low-hard states \citep{Remillard_McClintock-2006}. In future work we will investigate whether our conclusions about NTPA and radiative signatures still hold in the presence of ions and pair production. We will also study effects of the Comptonization (e.g., secondary scatterings) on the escaping radiation. This will allow first-principles prediction of the IC spectrum of flares powered by magnetic reconnection.

\section*{Acknowledgements}
We thank M.\,Begelman, A.\,Beloborodov, E.\,Quataert, K.\,Parfrey, L.\,Sironi, A.\,Spitkovsky, and V.\,Zhdankin for fruitful discussions. This work was supported by DOE grant DE-SC0008409, 
NASA grants NNX17AK57G and NNX16AB28G, NSF grant AST-1411879, and NASA through Einstein Postdoctoral Fellowship grant PF7-180165 awarded to AP by the Chandra X-ray Center, operated by the Smithsonian Astrophysical Observatory for NASA under contract NAS803060.  The simulations presented in this paper used computational resources of the NASA/Ames HEC Program and XSEDE/Stampede2 (allocation PHY140041) \citep{XSEDE2014}. DAU gratefully acknowledges the hospitality of the Inst. for Advanced Study and support from the Ambrose Monell Foundation.








\bsp	
\label{lastpage}
\end{document}